\documentclass[12pt]{iopart}
\usepackage{graphicx}

\begin{document}

\title[Generation of a coherent squeezed like state using a nonlinear photonic crystal]{Generation of a coherent squeezed like state defined with the Lie--Trotter product formula using a nonlinear photonic crystal}

\author{Hiroo Azuma}

\address{Global Research Center for Quantum Information Science,
National Institute of Informatics,
2-1-2 Hitotsubashi, Chiyoda-ku, Tokyo 101-8430, Japan}
\ead{hiroo.azuma@m3.dion.ne.jp, zuma@nii.ac.jp}
\vspace{10pt}
\begin{indented}
\item[]April 2023
\end{indented}

\begin{abstract}
In this paper, we investigate how to generate coherent squeezed like light using a nonlinear photonic crystal.
Because the photonic crystal reduces the group velocity of the incident light,
if it is composed of a material with a second-order nonlinear optical susceptibility $\chi^{(2)}$,
the interaction between the nonlinear material and the light passing through it strengthens and the quantum state of the emitted light is largely squeezed.
Thus, we can generate a coherent squeezed like light with a resonating cavity in which the nonlinear photonic crystal is placed.
This coherent squeezed like state is defined with the Lie--Trotter product formula and its mathematical expression is different from those of conventional coherent squeezed states.
We show that we can obtain this coherent squeezed like state with a squeezing level $15.9$ dB practically by adjusting physical parameters for our proposed method.
Feeding the squeezed light whose average number of photons is given by one or two into a beam splitter
and splitting the flow of the squeezed light into a pair of entangled light beams,
we estimate their entanglement quantitatively.
This paper is a sequel to
H.~Azuma, J. Phys. D: Appl. Phys. \textbf{55}, 315106 (2022).
\end{abstract}

%

%
\submitto{\JPD}
%
%
%

\section{\label{section-introduction}Introduction}
A photonic crystal is an optical device constructed periodically from parts of two materials whose refractive indexes are different from each other.
It causes an effect on the propagation of incident light and gives rise to a photonic band gap of conduction
so that the light with specific wavelengths cannot pass through it \cite{Yablonovitch1987,John1987,Yablonovitch1993}.
We can derive this dispersion relation for a one-dimensional photonic crystal with Maxwell's equations,
the Kronig--Penney model, and Bloch's theorem \cite{Kronig1931,Kittel1996}.
Thus, the photonic band gap of the photonic crystal is not a result of quantum mechanics but that of classical physics.
However, we can utilize photonic crystals to cause quantum effects on incident photons.
Because the photonic crystal reduces the group velocity of the photon passing through it,
it can make the time-of-flight of the photon longer.
Thus, if one of the materials that compose the photonic crystal has a nonlinear characteristic,
the duration of nonlinear interaction between the photon and the material increases and we can obtain a strong nonlinear effect.
Thus, we can regard the photonic crystal as an amplifier of the nonlinear interaction between the photon and the material.

A squeezed state is a quantum state of photons whose standard deviations of two operators,
$\hat{x}_{1}=(\hat{a}+\hat{a}^{\dagger})/2$ and $\hat{x}_{2}=(\hat{a}-\hat{a}^{\dagger})/(2i)$
where $\hat{a}$ and $\hat{a}^{\dagger}$ represent annihilation and creation operators of the photon,
satisfy
$\Delta x_{1}<1/2$ or $\Delta x_{2}<1/2$ with $\Delta x_{1}\Delta x_{2}\geq 1/4$
\cite{Walls1994,Barnett1997}.
In contrast, although a coherent state has minimal uncertainty $\Delta x_{1}\Delta x_{2}=1/4$,
the standard deviations are equal to each other,
$\Delta x_{1}=\Delta x_{2}=1/2$.
One of the nonlinear interactions that generate the squeezed state is the spontaneous parametric down conversion (SPDC) process in a nonlinear medium
with no inversion symmetry (or with nonzero $\chi^{(2)}$).
In general, many researchers consider that experimental generation of the squeezed state with a large squeezing level is difficult
and it is one of the most active topics in the field of quantum optics.
Historically, realization of strong squeezing was discussed using cavity degenerate-parametric-amplifier and cavity four-wave-mixing configurations \cite{Yurke1984},
parametric down conversion \cite{Wu1986}, and so on.
Research on this subject develops rapidly at present
because squeezed states play an important role in gravitational wave detection \cite{Eberle2010,Singh2019}.
For example, as a cutting-edge result, an experimental demonstration of the detection of a 15 dB squeezed state of light was reported in \cite{Vahlbruch2016}.

It is pointed out that a beam splitter splits the flow of the squeezed light into a pair of entangled light beams \cite{Kim2002,Azuma2022}.
Thus, many researchers consider that this phenomenon can be utilized for quantum information processing.
In particular, the generation of squeezed light with a simple small device has a marked academic impact on the field of quantum optics.
Hence, producing the squeezed light conveniently is a challenging task for the community of quantum information science.

In this work, we assume that the photonic crystal is composed of LiNbO${}_{3}$ and air.
Such photonic crystals are produced by plasma etching approaches (for example, argon-ion milling) and electron-beam lithography
in the laboratories \cite{Liang2017,Li2018,Li2020,Jiang2020}.
Because LiNbO${}_{3}$ has a large second-order nonlinear susceptibility $\chi^{(2)}$,
this photonic crystal transforms the incident coherent light into the squeezed coherent light \cite{Azuma2022,Sakoda2005}.
This can be a motivation for producing a one-dimensional photonic crystal from a material with nonlinear susceptibility $\chi^{(2)}$.
However, demonstrating this method actually in the laboratory is difficult because of the following two reasons:
\begin{enumerate}
\item We can consider BBO and $\mbox{LiNbO}_{3}$ as candidates for nonlinear materials.
However, for example, the nonlinear susceptibility of $\mbox{LiNbO}_{3}$ is given by
$\chi^{(2)}=\epsilon_{0}\tilde{\chi}^{(2)}$ and $\tilde{\chi}^{(2)}=25.2\times 10^{-12}$ mV${}^{-1}$
where $\epsilon_{0}$ represents the permittivity of vacuum \cite{Shoji1997,Kawase2002,Schiek2012}.
It is too small to obtain large squeezing such as $9.0$ dB.
\item Because the photonic crystal reduces the group velocity of the photon passing through it,
we can increase a nonlinear effect caused by $\chi^{(2)}$.
To realize this effect, we must make the group velocity of the photon $v_{\mbox{\scriptsize g}}$
be given by $v_{\mbox{\scriptsize g}}/c=4.57 \times 10^{-3}$
for obtaining squeezing level $8.69$ dB,
that is to say, squeezing parameter $r=1$, according to \cite{Azuma2022}.
Moreover, at the same time, we must adjust the frequency of the incident signal light with high precision,
$\nu_{\mbox{\scriptsize s}}=3.23 \times 10^{14}\pm 3.11\times 10^{8}$ Hz.
This requirement is very severe for actual experiments.
\end{enumerate}

In this work, to overcome the above two problems,
we propose that we can generate the coherent squeezed like light with a ring resonator where the one-dimensional nonlinear photonic crystal is placed.
Because the light circles around the ring resonator many times,
we can make the total time-of-flight of the photons in the photonic crystal longer.
Therefore, we can expect to obtain a strong squeezing level.

If we adopt this method, we need neither to reduce the group velocity of the light extremely nor to adjust the frequency of the light very precisely.
However, the state that we obtain by this method cannot be written in the form
of a conventional squeezed coherent state
$|\zeta,\alpha\rangle=S(\zeta)D(\alpha)|0\rangle$,
where $|0\rangle$, $D(\alpha)$, and $S(\zeta)$ represent the vacuum state, a displacement operator, and a squeezing operator, respectively,
but it is given by $[D(\alpha/N)S(\zeta/N)]^{N}|0\rangle$,
where $N$ represents the number of laps the light travels around the ring resonator.
We examine this state theoretically.

Further,
we show that we can obtain the squeezing level of $15.9$ dB by setting physical parameters properly.
Our result is comparable to state-of-the-art technologies,
for example, \cite{Vahlbruch2016}, in importance.
We derive an explicit form of entangled light beams emitted from a 50-50 beam splitter into which the squeezed light is injected
and estimate their entanglement quantitatively for $\alpha\sim 1$,
that is to say,
the average number of photons is nearly equal to one or two.

This paper is organized as follows.
In Sec.~\ref{section-setups-photonic-crystal-ring-resonator},
we explain the set-ups of the photonic crystal and the ring resonator that are used for our proposed method.
In Sec.~\ref{section-Lie-Trotter},
we examine relationships between the coherent squeezed like state defined with the Lie--Trotter product formula and the conventional squeezed coherent states.
In Sec.~\ref{section-uncertainty},
we derive an uncertainty relation of the coherent squeezed like state defined with the Lie--Totter product formula.
In Sec.~\ref{section-physical-parameters},
we numerically estimate the physical parameters of the photonic crystal and the ring resonator for generating the coherent squeezed like light.
In Sec.~\ref{section-entanglement-squeezed-state},
we quantitatively estimate entanglement between the light beams emitted from the photonic crystal and the 50-50 beam splitter.
In Sec.~\ref{section-discussions},
we give brief discussions.

\section{\label{section-setups-photonic-crystal-ring-resonator}Set-ups of the photonic crystal and the ring resonator}

\begin{figure}
\begin{center}
\includegraphics[scale=0.6]{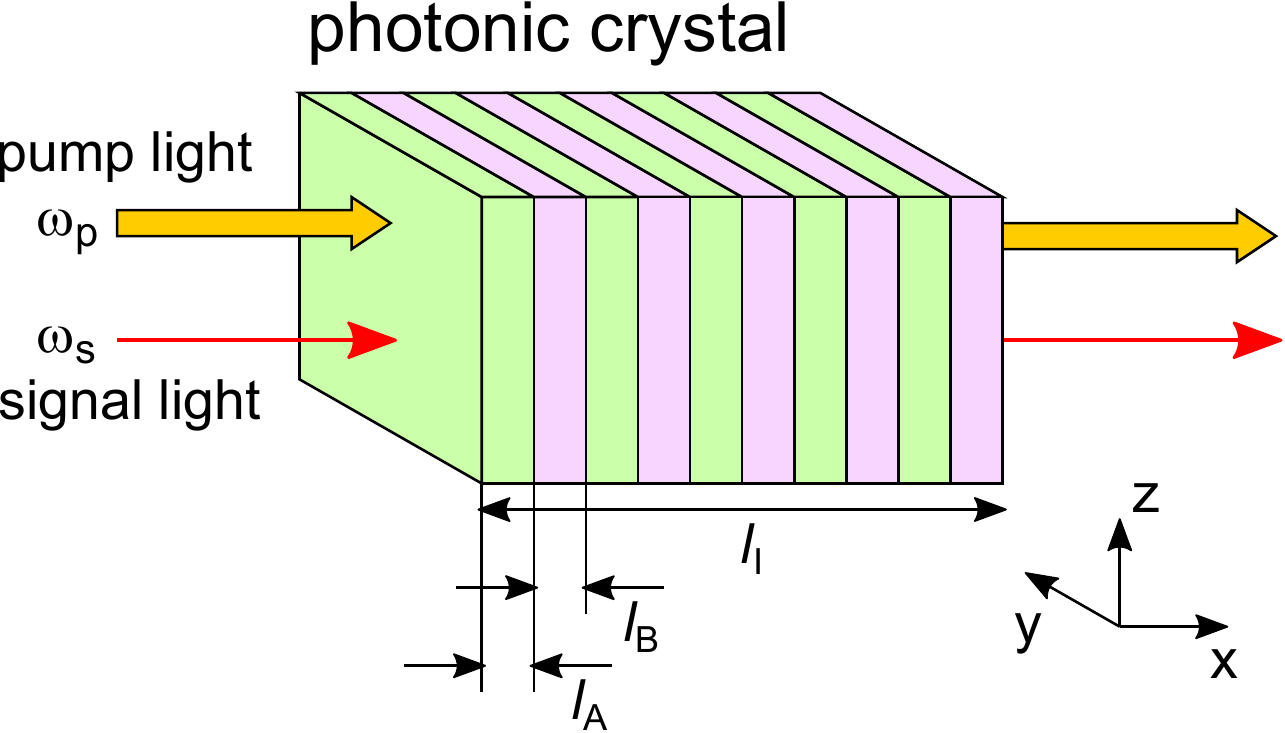}
\end{center}
\caption{Schematic of one-dimensional nonlinear photonic crystal,
and incident pump and signal light beams.
The materials A and B are LiNbO${}_{3}$ and air, respectively.}
\label{Figure01}
\end{figure}

\begin{figure}
\begin{center}
\includegraphics[scale=0.75]{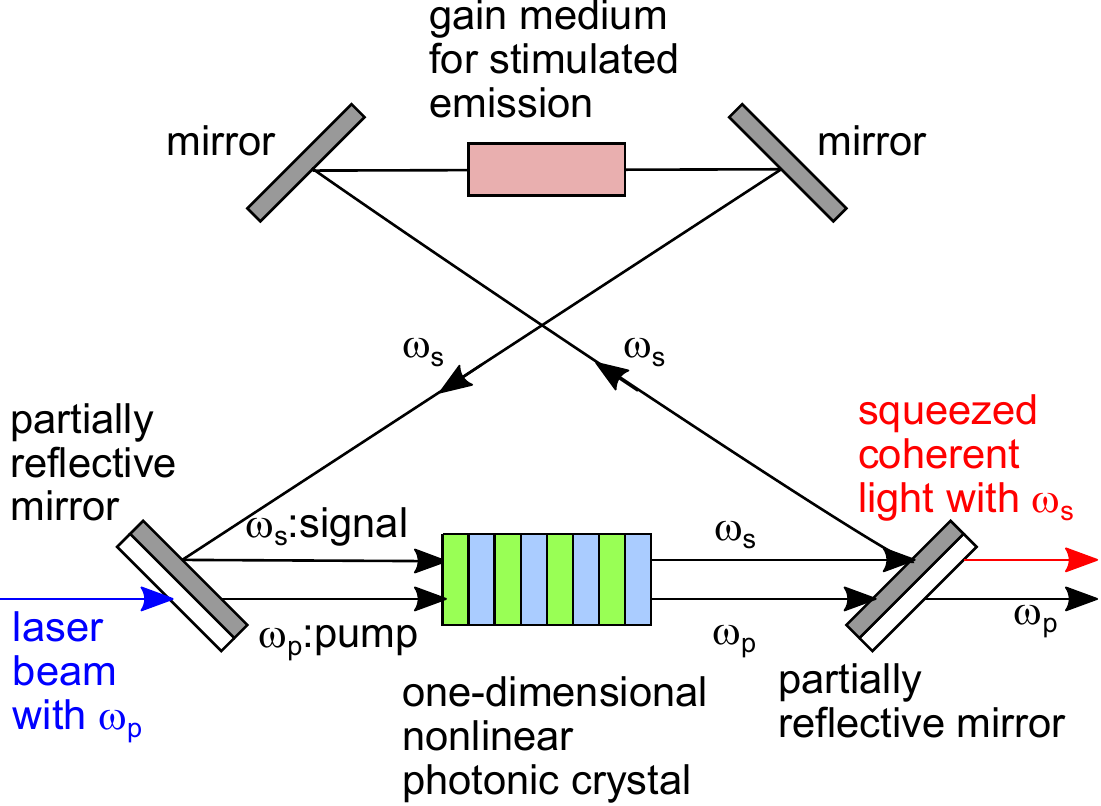}
\end{center}
\caption{Resonator implemented by bow-tie-type ring cavity.
In the resonator, the one-dimensional nonlinear photonic crystal and the gain medium for stimulated emission are placed.
The photonic crystal generates the squeezed state.
The gain medium generates coherent light.
The resonator outputs a coherent squeezed like light whose angular frequency is given by $\omega_{\mbox{\scriptsize s}}$.}
\label{Figure02}
\end{figure}

The one-dimensional photonic crystal is constructed from many layers of materials whose refractive indexes are different from each other.
We show its schematic in Fig.~\ref{Figure01}.
In this work, we assume that materials A and B are LiNbO${}_{3}$ and air.
We consider a bow-tie-type resonator shown in Fig.~\ref{Figure02} as a ring cavity.
Resonance is caused in the bow-tie-type ring resonator,
in which the signal radiation passes through and changes into coherent squeezed like light.
Details of functions of an optical circuit shown in Figs.~\ref{Figure01} and \ref{Figure02} are as follows \cite{Sakoda2005,Azuma2022}.

First, we explain Fig.~\ref{Figure01}.
We express the electric field of a pump light beam injected into the photonic crystal as $E_{\mbox{\scriptsize p}}$.
Because we assume that $E_{\mbox{\scriptsize p}}$ is strong enough, we can neglect its quantum fluctuation.
By contrast, an amplitude of a signal light beam fed into the photonic crystal is very weak,
so we need to consider it as a quantum one.
Moreover, we assume that angular frequencies of the signal and pump photons $\omega_{\mbox{\scriptsize p}}$ and $\omega_{\mbox{\scriptsize s}}$ satisfy
a condition $\omega_{\mbox{\scriptsize p}}=2\omega_{\mbox{\scriptsize s}}$.
We express wave vectors of the signal and pump photons as 
$\mbox{\boldmath $k$}_{\mbox{\scriptsize s}}=(k_{\mbox{\scriptsize s}},0,0)$ and
$\mbox{\boldmath $k$}_{\mbox{\scriptsize p}}=(k_{\mbox{\scriptsize p}},0,0)$
and assume that
$k_{\mbox{\scriptsize p}}
=
2k_{\mbox{\scriptsize s}}$
holds.
The electric fields of the pump and signal photons of
$\mbox{\boldmath $E$}_{\mbox{\scriptsize p}}
=
(0,E_{\mbox{\scriptsize p}},0)$
and
$\hat{\mbox{\boldmath $E$}}_{\mbox{\scriptsize s}}
=
(0,\hat{E}_{\mbox{\scriptsize s}},0)$
are described by the forms:
\begin{eqnarray}
E_{\mbox{\scriptsize p}}(x,t)
&=&
iA
\{
\exp[i(k_{\mbox{\scriptsize p}}x-\omega_{\mbox{\scriptsize p}}t+\theta)] \nonumber \\
&&
-
\exp[-i(k_{\mbox{\scriptsize p}}x-\omega_{\mbox{\scriptsize p}}t+\theta)]
\},
\end{eqnarray}
\begin{eqnarray}
\hat{E}_{\mbox{\scriptsize s}}(x,t)
&=&
i
\sqrt{\frac{\hbar\omega_{\mbox{\scriptsize s}}}{2\epsilon_{0}V}}
\{
\hat{a}_{\mbox{\scriptsize s}}\exp[i(k_{\mbox{\scriptsize s}}x-\omega_{\mbox{\scriptsize s}}t)] \nonumber \\
&&
-
\hat{a}_{\mbox{\scriptsize s}}^{\dagger}\exp[-i(k_{\mbox{\scriptsize s}}x-\omega_{\mbox{\scriptsize s}}t)]
\},
\label{electric-field-incident-signal-light-0}
\end{eqnarray}
where $\theta$, $V$, $A$, $\hat{a}_{\mbox{\scriptsize s}}$, and $\hat{a}^{\dagger}_{\mbox{\scriptsize s}}$
represent a phase of the pump light,
a volume for quantization of the signal light, an amplitude of the pump light,
and annihilation and creation operators of the signal photon, respectively.
In Eq.~(\ref{electric-field-incident-signal-light-0}), $\hbar$ is equal to $h/(2\pi)$ and $h$ represents the Planck constant.

In the above paragraph, we assume $\omega_{\mbox{\scriptsize p}}=2\omega_{\mbox{\scriptsize s}}$.
To prepare signal and pump laser beams whose angular frequencies satisfy this condition,
we can use techniques of second-harmonic generation with nonlinear medium.
In \cite{Guo2010}, an experimental demonstration of second-harmonic generation with PPLN (periodically poled lithium niobate)
and PPKTP (periodically poled KTP, potassium titanyl phosphate, $\mbox{KO}_{5}\mbox{PTi}$) crystals was reported.
In \cite{Eto2008}, for an experiment of observation of quadrature squeezing in a $\chi^{(2)}$ nonlinear waveguide,
the second-harmonic generation was performed with the PPLN-WG (PPLN-waveguide).
In these experiments, injecting a coherent light field with an angular frequency $\omega$ into a nonlinear optical medium with $\chi^{(2)}$,
one can obtain a coherent light field with an angular frequency $2\omega$.
This scheme is well-established,
and we can prepare the signal and pump photons that satisfy $\omega_{\mbox{\scriptsize p}}=2\omega_{\mbox{\scriptsize s}}$
rigorously with this method.

To generate the squeezed light with the photonic crystal,
we need to satisfy a phase-matching condition,
$\mbox{\boldmath $k$}_{\mbox{\scriptsize p}}
=
2\mbox{\boldmath $k$}_{\mbox{\scriptsize s}}
+
\mbox{\boldmath $G$}$,
where $\mbox{\boldmath $G$}$
represents an arbitrary reciprocal lattice vector.
Because we consider a one-dimensional photonic crystal,
$\mbox{\boldmath $G$}$ can be equal to zero or perpendicular to the long axis of the photonic crystal.
If we set $\mbox{\boldmath $G$}=\mbox{\boldmath $0$}$,
we can let $\mbox{\boldmath $k$}_{\mbox{\scriptsize p}}$ and $\mbox{\boldmath $k$}_{\mbox{\scriptsize s}}$
be parallel.
(The phase matching condition is given by Eq.~(1.26) in \cite{Sakoda2005}.
The definition of the reciprocal lattice vector is given in Chap.~2 of \cite{Kittel1996}
and Chap.~1 of \cite{Ziman1972}.
The one-dimensional photonic crystal has only one primitive transition vector and we write it as $\mbox{\boldmath $a$}$.
Because we can set $\mbox{\boldmath $G$}\cdot \mbox{\boldmath $a$}=0$
or $\mbox{\boldmath $G$}\cdot \mbox{\boldmath $a$}=2\pi$
and $\mbox{\boldmath $a$}$ is parallel to the long axis of the photonic crystal,
we can select $\mbox{\boldmath $G$}$ that is equal to zero or $2\pi\mbox{\boldmath $a$}/|\mbox{\boldmath $a$}|^{2}$,
or perpendicular to $\mbox{\boldmath $a$}$.)

We can express the signal light emitted from the photonic crystal whose coordinate of the position is given by $x$ as
\begin{eqnarray}
\hat{E}_{\mbox{\scriptsize s}}(x,t)
&=&
i
\sqrt{\frac{\hbar\omega_{\mbox{\scriptsize s}}}{2\epsilon_{0}V}}
\{
\hat{b}_{\mbox{\scriptsize s}}(x)\exp[i(k_{\mbox{\scriptsize s}}x-\omega_{\mbox{\scriptsize s}}t)] \nonumber \\
&&
-
\hat{b}^{\dagger}_{\mbox{\scriptsize s}}(x)\exp[-i(k_{\mbox{\scriptsize s}}x-\omega_{\mbox{\scriptsize s}}t)]
\},
\end{eqnarray}
\begin{equation}
\hat{b}_{\mbox{\scriptsize s}}(x)
=
\hat{a}_{\mbox{\scriptsize s}}\cosh(|\beta| x)
+
\hat{a}_{\mbox{\scriptsize s}}^{\dagger}\exp[i(\theta+\phi)]\sinh(|\beta| x),
\end{equation}
\begin{equation}
\beta
=
\frac{\omega_{\mbox{\scriptsize s}}A\chi^{(2)}}{\epsilon_{0}v_{\mbox{\scriptsize g}}},
\label{photonic-crystal-squeezing-para-01}
\end{equation}
where a phase $\phi$ is defined as $e^{i\phi}=\beta/|\beta|$ and $v_{\mbox{\scriptsize g}}$ is a group velocity of the signal photon passing through the photonic crystal.
An operator $\hat{b}_{\mbox{\scriptsize s}}(x)$ is obtained by the Bogoliubov transform of $\hat{a}_{\mbox{\scriptsize s}}$.

Next, defining a squeezing operator as
\begin{equation}
\hat{S}(\zeta)
=
\exp(
-\frac{\zeta}{2}\hat{a}_{\mbox{\scriptsize s}}^{\dagger 2}
+\frac{\zeta^{*}}{2}\hat{a}_{\mbox{\scriptsize s}}^{2}
),
\label{single-mode-squeezing-operator-01}
\end{equation}
\begin{equation}
\zeta
=
\beta x \exp(i\theta)
=
|\beta|x \exp[i(\theta+\phi)],
\label{squeezing-parameter-zeta-1}
\end{equation}
we obtain the following relation:
\begin{equation}
\hat{S}(\zeta)\hat{a}_{\mbox{\scriptsize s}}\hat{S}^{\dagger}(\zeta)
=
\hat{b}_{\mbox{\scriptsize s}}(x).
\end{equation}
Thus, if we inject a coherent light $|\alpha\rangle$ into the photonic crystal,
a squeezed coherent light $|-\zeta,\alpha\rangle$ is emitted from it in the form,
\begin{eqnarray}
|-\zeta,\alpha\rangle
&=&
\hat{S}^{\dagger}(\zeta)|\alpha\rangle \nonumber \\
&=&
\hat{S}(-\zeta)|\alpha\rangle.
\label{squeezed-coherent-state-emitted-photonic-crystal-1}
\end{eqnarray}
From now on, in Secs.~\ref{section-setups-photonic-crystal-ring-resonator},
\ref{section-physical-parameters}, and \ref{section-entanglement-squeezed-state},
for the sake of simplicity,
we assume $\theta=\phi=0$
and we describe the squeezing parameter $\zeta$ defined in Eq.~(\ref{squeezing-parameter-zeta-1}) as a real value $r$,
that is to say, $r=|\beta|l_{\mbox{\scriptsize I}}$,
where $l_{\mbox{\scriptsize I}}$ represents the width of the photonic crystal.
Further, we assume that the parameter of the coherent light $\alpha$ is real.

Next, we explain Fig.~\ref{Figure02}.
After passing through the nonlinear photonic crystal,
the signal light beam goes around the ring resonator many times and develops into a coherent squeezed like light.
We assume that the beam goes around the ring resonator $N$ times,
is transmitted through a partially reflective mirror,
and flies out of the resonator.
While the signal light passes through the photonic crystal with the time-of-flight $\Delta\tau_{1}$,
the following Hamiltonian is induced:
\begin{equation}
\hat{H}_{\mbox{\scriptsize I}}
=
-i\frac{\kappa}{2}
(\hat{a}_{\mbox{\scriptsize s}}^{\dagger 2}-\hat{a}_{\mbox{\scriptsize s}}^{2}),
\end{equation}
where
\begin{equation}
\kappa
=
\frac{\hbar}{\Delta\tau_{1}}\beta l_{\mbox{\scriptsize I}}.
\end{equation}
Remembering that the length of the photonic crystal and the group velocity of the signal light are given
by $l_{\mbox{\scriptsize I}}$ and $v_{\mbox{\scriptsize g}}$, respectively,
we obtain
$l_{\mbox{\scriptsize I}}=v_{\mbox{\scriptsize g}}\Delta\tau_{1}$.
Thus, we arrive at
\begin{equation}
\kappa
=
\frac{\hbar\omega_{\mbox{\scriptsize s}}A\chi^{(2)}}{\epsilon_{0}}.
\end{equation}

On the other hand, while the signal light passes through a gain medium for stimulated emission,
the following Hamiltonian is induced:
\begin{equation}
\hat{H}_{\mbox{\scriptsize II}}
=
i\hbar
(\gamma \hat{a}_{\mbox{\scriptsize s}}^{\dagger}
-
\gamma^{*}\hat{a}_{\mbox{\scriptsize s}}).
\label{Hamiltonian-II-displacement-operator-0}
\end{equation}
The above Hamiltonian was mentioned in Eqs.~(9.19), (9.20), and (9.21) in \cite{Glauber1963}.
A unitary transformation caused by the Hamiltonian $\hat{H}_{\mbox{\scriptsize II}}$ corresponds to the displacement operator,
which is given by
\begin{equation}
\hat{T}(l_{\mbox{\scriptsize II}})
=
\exp(\frac{l_{\mbox{\scriptsize II}}}{i\hbar}\hat{p}_{\mbox{\scriptsize s}}),
\end{equation}
\begin{equation}
\hat{p}_{\mbox{\scriptsize s}}
=
i
\sqrt{\frac{m_{\mbox{\scriptsize s}}\omega_{\mbox{\scriptsize s}}\hbar}{2}}
(\hat{a}_{\mbox{\scriptsize s}}^{\dagger}-\hat{a}_{\mbox{\scriptsize s}}),
\end{equation}
where $m_{\mbox{\scriptsize s}}$ denotes the virtual mass of the signal photon regarded as a harmonic oscillator
and $\hat{p}_{\mbox{\scriptsize s}}$ represents the momentum operator of the signal photon.
The length $l_{\mbox{\scriptsize II}}$ is a displacement added to the position operator $\hat{x}_{\mbox{\scriptsize s}}$.

Here, we draw attention to the fact
$m_{\mbox{\scriptsize s}}\omega_{\mbox{\scriptsize s}}
=2\pi m_{\mbox{\scriptsize s}}\nu_{\mbox{\scriptsize s}}
=2\pi p_{\mbox{\scriptsize s}}/\lambda_{\mbox{\scriptsize s}}$,
where $\nu_{\mbox{\scriptsize s}}$, $p_{\mbox{\scriptsize s}}$, and $\lambda_{\mbox{\scriptsize s}}$
represent the frequency, momentum, and wavelength of the signal photon, respectively.
Thus, we can derive a relationship
$m_{\mbox{\scriptsize s}}\omega_{\mbox{\scriptsize s}}\hbar/2
=\pi p_{\mbox{\scriptsize s}}\hbar/\lambda_{\mbox{\scriptsize s}}
=\pi h\hbar/\lambda_{\mbox{\scriptsize s}}^{2}
=(2\pi\hbar)^{2}/(2\lambda_{\mbox{\scriptsize s}}^{2})$,
where we use $p_{\mbox{\scriptsize s}}=h/\lambda_{\mbox{\scriptsize s}}$.
Hence, we obtain
\begin{equation}
\hat{T}(l_{\mbox{\scriptsize II}})
=
\exp
[
\sqrt{2}\pi\frac{l_{\mbox{\scriptsize II}}}{\lambda_{\mbox{\scriptsize s}}}
(\hat{a}_{\mbox{\scriptsize s}}^{\dagger}-\hat{a}_{\mbox{\scriptsize s}})
].
\end{equation}
Describing $\hat{T}(l_{\mbox{\scriptsize II}})$ in the form:
\begin{equation}
\hat{T}(l_{\mbox{\scriptsize II}})
=
\exp
[
-i\frac{\Delta\tau_{2}}{\hbar}\hat{H}_{\mbox{\scriptsize II}}
],
\end{equation}
and regarding $l_{\mbox{\scriptsize II}}$ as a length of the gain medium,
we obtain
\begin{equation}
\sqrt{2}\pi\frac{l_{\mbox{\scriptsize II}}}{\lambda_{\mbox{\scriptsize s}}}
=
\Delta\tau_{2}\gamma,
\end{equation}
where we assume that $\gamma$ is real.
Because $l_{\mbox{\scriptsize II}}=c\Delta\tau_{2}$, we reach
\begin{equation}
\gamma
=
\sqrt{2}\pi\frac{c}{\lambda_{\mbox{\scriptsize s}}}.
\label{gamma-ideal-1}
\end{equation}

The above estimation is correct if all of the photons injected into the gain medium contribute to the generation of the coherent light
and every photon in the flux of the signal beam occupies a volume $\sigma\lambda_{\mbox{\scriptsize s}}$
where $\sigma$ represents a cross-section of the gain medium.
However, to evaluate $\gamma$ practically,
we need to consider a cross-section and a density of excited atoms in the gain medium.
Thus, instead of Eq.~(\ref{gamma-ideal-1}), we should describe $\gamma$ as
\begin{equation}
\gamma
=
\sqrt{2}\pi\frac{c}{\lambda_{\mbox{\scriptsize s}}}
\frac{\sigma_{\mbox{\scriptsize em}}}{\sigma}\rho_{0}\Delta V,
\label{gamma-practical-2}
\end{equation}
where $\sigma_{\mbox{\scriptsize em}}$, $\rho_{0}$, and $\Delta V$
denote the stimulated emission cross-section for the excited atom, the density of the excited atoms in the gain medium, and the volume of the gain medium, respectively.

Assuming that times-of-flight for the photon to pass through the photonic crystal and the gain medium are given by $\Delta\tau_{1}$ and $\Delta\tau_{2}$,
respectively,
we can write down the state of the signal light emitted from the resonator as
\begin{equation}
\Bigl(
\exp[-i\frac{\Delta\tau_{2}}{\hbar}\hat{H}_{\mbox{\scriptsize II}}]
\exp[-i\frac{\Delta\tau_{1}}{\hbar}\hat{H}_{\mbox{\scriptsize I}}]
\Bigr)^{N}
|0\rangle.
\label{Lie-Trotter-squeezed-coherent-state-1}
\end{equation}
Now, setting $T^{(1)}=N\Delta\tau_{1}$ and $T^{(2)}=N\Delta\tau_{2}$,
we can rewrite the above equation as
\begin{equation}
\Bigl(
\exp[-i\frac{T^{(2)}}{N\hbar}\hat{H}_{\mbox{\scriptsize II}}]
\exp[-i\frac{T^{(1)}}{N\hbar}\hat{H}_{\mbox{\scriptsize I}}]
\Bigr)^{N}
|0\rangle.
\label{Lie-Trotter-squeezed-coherent-state-2}
\end{equation}

\section{\label{section-Lie-Trotter}A coherent squeezed like state defined with the Lie--Trotter product formula}
So far, there are conventional squeezed coherent and coherent squeezed states,
\begin{eqnarray}
|\zeta,\alpha\rangle
&=&
\hat{S}(\zeta)\hat{D}(\alpha)|0\rangle, \nonumber \\
||\zeta,\alpha\rangle\!\rangle
&=&
\hat{D}(\alpha)\hat{S}(\zeta)|0\rangle,
\label{conventional-squeezed-coherent-states-0}
\end{eqnarray}
where
\begin{eqnarray}
\hat{D}(\alpha)
&=&
\exp(\alpha \hat{a}^{\dagger}-\alpha^{*} \hat{a}), \nonumber \\
\hat{S}(\zeta)
&=&
\exp[\frac{1}{2}(\zeta^{*}\hat{a}^{2}-\zeta \hat{a}^{\dagger 2})].
\end{eqnarray}
In Secs.~\ref{section-Lie-Trotter} and \ref{section-uncertainty},
we assume that $\alpha$ and $\zeta$ are complex numbers and $\zeta$ is written in the form $\zeta=r e^{i\varphi}$.

In this work, drawing attention to Eqs.~(\ref{Lie-Trotter-squeezed-coherent-state-1}) and (\ref{Lie-Trotter-squeezed-coherent-state-2}),
we propose the third state as follows:
\begin{equation}
|||\zeta,\alpha\rangle\!\rangle\!\rangle
=
\lim_{N\to\infty}
[
\hat{D}(\alpha)^{1/N}
\hat{S}(\zeta)^{1/N}
]^{N}
|0\rangle.
\end{equation}
Then, we use the Lie--Trotter product formula,
\begin{equation}
\lim_{N\to\infty}
(e^{it\hat{A}/N}e^{it\hat{B}/N})^{N}
=
e^{it(\hat{A}+\hat{B})}
\quad
\forall t>0,
\label{Lie-Trotter-formula-N-limit-0}
\end{equation}
where $\hat{A}$ and $\hat{B}$ represent arbitrary Hermitian operators
\cite{Trotter1959,Kato1978,Reed1980,Azuma2015}.
Thus, we obtain
\begin{equation}
|||\zeta,\alpha\rangle\!\rangle\!\rangle
=
\exp[\alpha\hat{a}^{\dagger}-\alpha^{*}\hat{a}
+\frac{1}{2}
(\zeta^{*}\hat{a}^{2}-\zeta\hat{a}^{\dagger 2})
]
|0\rangle.
\end{equation}

Now, we examine relationships between
$|\zeta,\alpha\rangle$, $||\zeta,\alpha\rangle\!\rangle$, and $|||\zeta,\alpha\rangle\!\rangle\!\rangle$.
Because of
\begin{equation}
\hat{S}(\zeta)\hat{D}(\alpha)\hat{S}^{\dagger}(\zeta)
=
\hat{D}(\alpha\cosh r-\alpha^{*}e^{i\varphi}\sinh r),
\label{formula-sS-D-S-dagger-0}
\end{equation}
we obtain the relationship between $|\zeta,\alpha\rangle$ and $||\zeta,\alpha\rangle\!\rangle$ as
\begin{eqnarray}
|\zeta,\alpha\rangle
&=&
\hat{S}(\zeta)\hat{D}(\alpha)\hat{S}^{\dagger}(\zeta)\hat{S}(\zeta)|0\rangle \nonumber \\
&=&
||\zeta,\alpha\cosh r-\alpha^{*}e^{i\varphi}\sinh r\rangle\!\rangle.
\label{conversion-squeezed-coherent-state-1}
\end{eqnarray}
Carrying out slightly tough calculations,
we obtain
\begin{eqnarray}
&&
[
\hat{D}(\alpha/N)
\hat{S}(\zeta/N)
]^{n} \nonumber \\
&=&
\exp
\Bigl\{
-i
\mbox{Im}[(\frac{\alpha}{N})^{2}e^{-i\varphi}] \nonumber \\
&&
\times
\Bigl[
\sum_{k=1}^{n-1}\sinh\frac{kr}{N}\sum_{m=1}^{k}\cosh\frac{(m-1)r}{N} \nonumber \\
&&
-
\sum_{k=1}^{n-1}\cosh\frac{kr}{N}\sum_{m=1}^{k}\sinh\frac{(m-1)r}{N}
\Bigr]
\Bigr\} \nonumber \\
&&
\times
\hat{D}(\frac{\alpha}{N}\sum_{k=1}^{n}\cosh\frac{(k-1)r}{N} \nonumber \\
&&
-
\frac{\alpha^{*}}{N}e^{i\varphi}\sum_{k=1}^{n}\sinh\frac{(k-1)r}{N}
)
\hat{S}(n\zeta/N) \nonumber \\
&&
\quad
\mbox{for $n=1,2,...,N-1$},
\end{eqnarray}
\begin{eqnarray}
&&
[
\hat{D}(\alpha/N)
\hat{S}(\zeta/N)
]^{N} \nonumber \\
&=&
\exp
\Bigl\{
-i
\mbox{Im}[(\frac{\alpha}{N})^{2}e^{-i\varphi}]
\frac{1}{4}{\rm{cosech}}^{2}(\frac{r}{2N}) \nonumber \\
&&
\times
(\sinh r-N\sinh(\frac{r}{N}))
\Bigr\} \nonumber \\
&&
\times
\hat{D}(\frac{\alpha}{2N}(1-\cosh r+\coth(\frac{r}{2N})\sinh r) \nonumber \\
&&
-
\frac{\alpha^{*}}{2N}e^{i\varphi}[(-1+\cosh r)\coth(\frac{r}{2N})-\sinh r])\hat{S}(\zeta),
\end{eqnarray}
\begin{eqnarray}
&&
\lim_{N\to\infty}[
\hat{D}(\alpha/N)
\hat{S}(\zeta/N)
]^{N} \nonumber \\
&=&
\exp
\Bigl[
-i\mbox{Im}[\alpha^{2}e^{-i\varphi}]
(
\frac{\sinh r}{r^{2}}-\frac{1}{r}
)
\Bigr] \nonumber \\
&&
\times
\hat{D}(\alpha\frac{\sinh r}{r}
-
\alpha^{*}e^{i\varphi}\frac{1}{r}(-1+\cosh r))\hat{S}(\zeta) \nonumber \\
&=&
\exp
\Bigl[
-i\mbox{Im}[\alpha^{2}e^{-i\varphi}]
(
\frac{\sinh r}{r^{2}}-\frac{1}{r}
)
\Bigr]
\hat{S}(\zeta) \nonumber \\
&&
\times
\hat{D}(\alpha\frac{\sinh r}{r}
+
\alpha^{*}e^{i\varphi}\frac{1}{r}(-1+\cosh r)).
\label{formula-S-zeta-D-alpha-N-Lie-Trotter-1}
\end{eqnarray}
Further, because of Eq.~(\ref{Lie-Trotter-formula-N-limit-0}),
we obtain
\begin{equation}
\lim_{N\to\infty}[
\hat{D}(\alpha/N)
\hat{S}(\zeta/N)
]^{N}
=
\lim_{N\to\infty}[
\hat{S}(\zeta/N)
\hat{D}(\alpha/N)
]^{N}.
\end{equation}
From the above results, we can derive the relationship between $|\zeta,\alpha\rangle$ and $|||\zeta,\alpha\rangle\!\rangle\!\rangle$ as
\begin{equation}
|||\zeta,\alpha\rangle\!\rangle\!\rangle
=
e^{i\Theta}
|\zeta,\tilde{\alpha}\rangle,
\label{squeezed-coherent-state-Lie-Trotter-1}
\end{equation}
\begin{equation}
\Theta
=
-\mbox{Im}[\alpha^{2}e^{i\varphi}]\frac{1}{r}
(\frac{\sinh r}{r}-1),
\label{squeezed-coherent-state-Lie-Trotter-2}
\end{equation}
\begin{equation}
\tilde{\alpha}
=
\alpha\frac{\sinh r}{r}
+
\alpha^{*}e^{i\varphi}\frac{1}{r}(-1+\cosh r).
\label{squeezed-coherent-state-Lie-Trotter-3}
\end{equation}

\section{\label{section-uncertainty}An uncertainty relation for $|||\zeta,\alpha\rangle\!\rangle\!\rangle$}
We derive an uncertainty relation between momentum and position operators for $|||\zeta,\alpha\rangle\!\rangle\!\rangle$.
We define the characteristic function of $|||\zeta,\alpha\rangle\!\rangle\!\rangle$ as follows:
\begin{equation}
\langle\!\langle\!\langle\mbox{CF}\rangle\!\rangle\!\rangle
=
\langle\!\langle\!\langle\zeta,\alpha|||
\exp[-i(q\hat{Q}+p\hat{P})]|||\zeta,\alpha\rangle\!\rangle\!\rangle,
\end{equation}
\begin{eqnarray}
\hat{P}
&=&
i
\sqrt{\frac{m\omega\hbar}{2}}(\hat{a}^{\dagger}-\hat{a}), \nonumber \\
\hat{Q}
&=&
\sqrt{\frac{\hbar}{2m\omega}}(\hat{a}^{\dagger}+\hat{a}),
\end{eqnarray}
\begin{equation}
-i(q\hat{Q}+p\hat{P})
=
\gamma \hat{a}^{\dagger}-\gamma^{*}\hat{a},
\end{equation}
\begin{equation}
\gamma
=
-iq\sqrt{\frac{\hbar}{2m\omega}}+p\sqrt{\frac{m\omega\hbar}{2}},
\end{equation}
where $p$ and $q$ are real values.
Dimensions of $p$ and $q$ are given by reciprocals of the momentum and position, respectively.
We regard the bosons represented by $\hat{a}$ and $\hat{a}^{\dagger}$ as a virtual harmonic oscillator whose mass and angular frequency are given
by $m$ and $\omega$, respectively.

The explicit form of $\langle\!\langle\!\langle\mbox{CF}\rangle\!\rangle\!\rangle$ is given by
\begin{equation}
\langle\!\langle\!\langle\mbox{CF}\rangle\!\rangle\!\rangle
=
\exp[-\frac{1}{2}|\kappa|^{2}+f(\alpha,\gamma,\zeta)],
\end{equation}
where
\begin{equation}
\kappa
=
\gamma\cosh r
+
\frac{\sinh r}{r}\zeta\gamma^{*},
\end{equation}
\begin{eqnarray}
f(\alpha,\gamma,\zeta)
&=&
\frac{\cosh r-1}{r^{2}}
(-\alpha\gamma\zeta^{*}+\alpha^{*}\gamma^{*}\zeta) \nonumber \\
&&
-
\frac{\sinh r}{r}(\alpha\gamma^{*}-\alpha^{*}\gamma).
\end{eqnarray}
Utilizing
$\langle\!\langle\!\langle\mbox{CF}\rangle\!\rangle\!\rangle$,
we can obtain expectation values of $\hat{Q}$, $\hat{Q}^{2}$, $\hat{P}$, and $\hat{P}^{2}$ straightforwardly.
For example, we can obtain the expectation value of $\hat{Q}$ by calculating
$
\langle\!\langle\!\langle Q\rangle\!\rangle\!\rangle
=
i(\partial/\partial q)\langle\!\langle\!\langle\mbox{CF}\rangle\!\rangle\!\rangle
|_{q=p=0}
$.
We can derive standard deviations as
\begin{eqnarray}
(\Delta Q)^{2}
&=&
\langle\!\langle\!\langle Q^{2}\rangle\!\rangle\!\rangle-\langle\!\langle\!\langle Q\rangle\!\rangle\!\rangle^{2} \nonumber \\
&=&
\frac{\hbar}{2m\omega}
\left|
\frac{\sinh r}{r}\zeta-\cosh r
\right|^{2},
\label{standard-deviation-Q-square}
\end{eqnarray}
\begin{eqnarray}
(\Delta P)^{2}
&=&
\langle\!\langle\!\langle P^{2}\rangle\!\rangle\!\rangle-\langle\!\langle\!\langle P\rangle\!\rangle\!\rangle^{2} \nonumber \\
&=&
\frac{m\omega\hbar}{2}
\left|
\frac{\sinh r}{r}\zeta+\cosh r
\right|^{2},
\label{standard-deviation-P-square}
\end{eqnarray}
\begin{eqnarray}
\Delta Q\Delta P
&=&
\frac{\hbar}{2}
\Bigl[
\sinh^{4}r+\cosh^{4}r \nonumber \\
&&
-
\frac{\zeta^{2}+\zeta^{*2}}{r^{2}}\sinh^{2}r\cosh^{2}r
\Bigr]^{1/2}.
\label{standard-deviation-Q-P}
\end{eqnarray}
Looking at Eqs.~(\ref{standard-deviation-Q-square}), (\ref{standard-deviation-P-square}), and (\ref{standard-deviation-Q-P}),
we note that $(\Delta Q)^{2}$, $(\Delta P)^{2}$, and $\Delta Q\Delta P$ do not depend on $\alpha$.
When $\zeta$ is real, that is to say,
$\zeta=\zeta^{*}=r$,
$\Delta Q\Delta P=\hbar/2$ holds with
\begin{eqnarray}
\Delta Q
&=&
\sqrt{\frac{\hbar}{2m\omega}}e^{-r}, \nonumber \\
\Delta P
&=&
\sqrt{\frac{m\omega\hbar}{2}}e^{r}.
\end{eqnarray}

\section{\label{section-physical-parameters}Physical parameters of the photonic crystal and the ring resonator for generating a coherent squeezed like light}
First, we estimate how large the squeezing level can be for actual experiments.
The LiNbO${}_{3}$ is a typical material that has a large second-order nonlinear optical susceptibility $\chi^{(2)}$.
The LiNbO${}_{3}$ is transparent for the light beam whose wavelength is between
$\lambda=4\times 10^{-7}$ m and $\lambda=4.8\times 10^{-6}$ m.
Its refractive index is given by $n\simeq 2.22$
\cite{Jundt1997,Gayer2008}.

In Fig.~\ref{Figure01},
we choose LiNbO${}_{3}$ and air for materials A and B, respectively.
Relative permittivities of LiNbO${}_{3}$ and air are given by
$\epsilon_{\mbox{\scriptsize A}}/\epsilon_{0}=n_{\mbox{\scriptsize A}}^{2}$ where $n_{\mbox{\scriptsize A}}=2.22$
and
$\epsilon_{\mbox{\scriptsize B}}/\epsilon_{0}=1$, respectively.
According to \cite{Liang2017,Li2018,Li2020,Jiang2020,Ishikawa2008},
we set $l_{\mbox{\scriptsize A}}=l_{\mbox{\scriptsize B}}=5.5 \times 10^{-7}$ m.

\begin{figure}
\begin{center}
\includegraphics[scale=0.9]{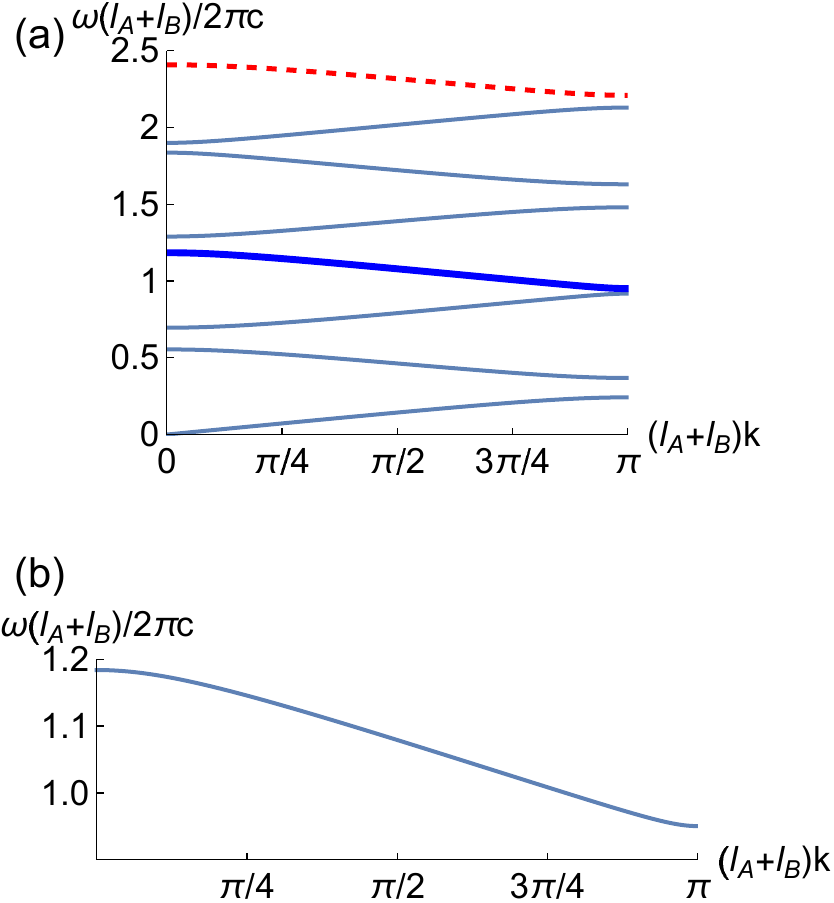}
\end{center}
\caption{(a) The dispersion relation of the photon passing through the photonic crystal that is composed of LiNbO${}_{3}$ and air.
The thick curve and the red dashed curve represent the fourth and eighth conduction bands from the bottom, respectively.
(b) The dispersion relation of the fourth conduction band from the bottom of the photonic crystal.}
\label{Figure03ab}
\end{figure}

\begin{figure}
\begin{center}
\includegraphics[scale=0.9]{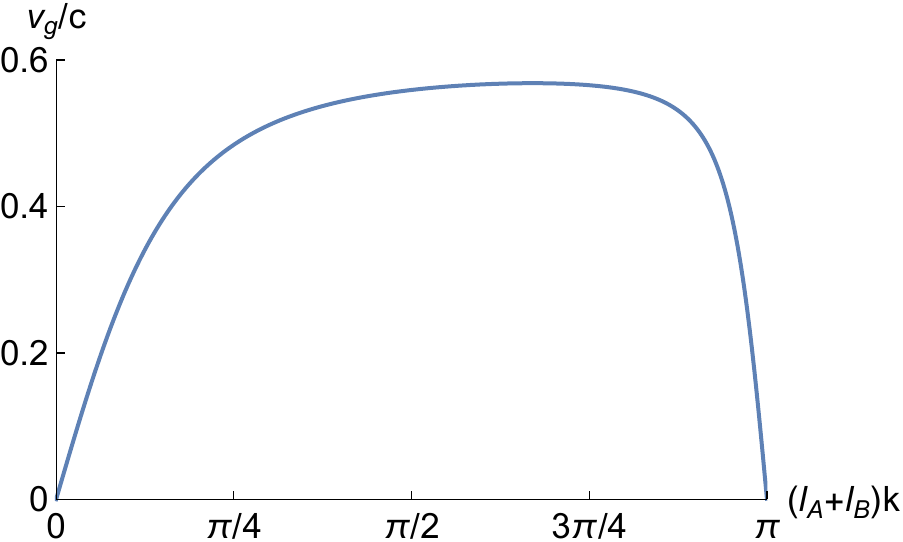}
\end{center}
\caption{A plot of the group velocity of the photon passing through the photonic crystal for the fourth conduction band
as a function of the wavenumber.}
\label{Figure04}
\end{figure}

Describing the wavenumber and angular frequency of the light flying in the photonic crystal as $k$ and $\omega=\omega(k)$,
we can give its dispersion relation as follows \cite{Azuma2008}:
\begin{eqnarray}
&&
\cos[(l_{\mbox{\scriptsize A}}+l_{\mbox{\scriptsize B}})k]
-
\cos(l_{\mbox{\scriptsize A}}K_{\mbox{\scriptsize A}})
\cos(l_{\mbox{\scriptsize B}}K_{\mbox{\scriptsize B}}) \nonumber \\
&&
+
\frac{K_{\mbox{\scriptsize A}}^{2}+K_{\mbox{\scriptsize B}}^{2}}{2K_{\mbox{\scriptsize A}}K_{\mbox{\scriptsize B}}}
\sin(l_{\mbox{\scriptsize A}}K_{\mbox{\scriptsize A}})
\sin(l_{\mbox{\scriptsize B}}K_{\mbox{\scriptsize B}})
=
0,
\label{dispersion-relation-2}
\end{eqnarray}
\begin{equation}
K_{\mbox{\scriptsize A}}
=
\frac{\omega}{c}\sqrt{\frac{\epsilon_{\mbox{\scriptsize A}}}{\epsilon_{0}}},
\label{dispersion-relation-1}
\quad
K_{\mbox{\scriptsize B}}
=
\frac{\omega}{c}\sqrt{\frac{\epsilon_{\mbox{\scriptsize B}}}{\epsilon_{0}}}.
\end{equation}
We plot this dispersion relation in Fig.~\ref{Figure03ab} (a).
Figure~\ref{Figure03ab} (a) shows eight conduction bands from the bottom.
Figure~\ref{Figure03ab} (b) shows the fourth conduction band from the bottom.
The group velocity of the light whose wavenumber is equal to $k_{0}$ in the photonic crystal is given by:
\begin{equation}
v_{\mbox{\scriptsize g}}(k_{0})
=
\left|
\frac{d\omega}{dk}(k_{0})
\right|.
\end{equation}
In Fig.~\ref{Figure04}, we plot the group velocity for the fourth conduction band from the bottom as a function of the wavenumber $k$.

Next, we evaluate a squeezing level obtained with our method practically.
The unitary transformation caused by the Hamiltonian $H_{\mbox{\scriptsize I}}$ is written down as
\begin{equation}
\exp[-i\frac{\Delta\tau_{1}}{\hbar}H_{\mbox{\scriptsize I}}]
=
\exp[-\frac{\Delta\tau_{1}}{2\hbar}\kappa(\hat{a}_{\mbox{\scriptsize s}}^{\dagger 2}-\hat{a}_{\mbox{\scriptsize s}}^{2})].
\end{equation}
Here, we compute the squeezing level of the light going around the ring resonator $N$ times.
Assuming that the squeezing parameter $\zeta$ defined in Eq.~(\ref{squeezing-parameter-zeta-1}) is real and expressing it as $r$,
we obtain
\begin{equation}
r
=
\frac{\Delta\tau_{1}}{\hbar}N\kappa
=
\frac{T^{(1)}}{\hbar}\frac{\hbar\omega_{\mbox{\scriptsize s}}A\chi^{(2)}}{\epsilon_{0}}
=
T^{(1)}\omega_{\mbox{\scriptsize s}}A\tilde{\chi}^{(2)},
\label{squeezing-parameter-r-1}
\end{equation}
where $\chi^{(2)}=\epsilon_{0}\tilde{\chi}^{(2)}$.

Now, we consider a scenario where the width of the layers of the photonic crystal is given by
$l_{\mbox{\scriptsize A}}=l_{\mbox{\scriptsize B}}=5.5 \times 10^{-7}$ m
and it consists of $20$ layers in total.
Thus, LiNbO${}_{3}$ and air are constructed from $10$ layers each,
so that $l_{\mbox{\scriptsize I}}=5.5\times 10^{-6}$ m.
The typical number of times the light goes around the ring resonator is given by $N=1000$ \cite{Endo2022}.
The typical group velocity of the signal light flying in the photonic crystal is given by $v_{\mbox{\scriptsize g}}/c=0.05$ \cite{Notomi2001,Asano2004}.
Thus,
because of $l_{\mbox{\scriptsize I}}=v_{\mbox{\scriptsize g}}\Delta\tau_{1}$ and $T^{(1)}=N\Delta\tau_{1}$,
we obtain $T^{(1)}=3.67\times 10^{-10}$ s.

Looking at Fig.~\ref{Figure04}, we obtain
$(l_{\mbox{\scriptsize A}}+l_{\mbox{\scriptsize B}})k_{\mbox{\scriptsize s}}=4.66\times 10^{-2}$
and
$k_{\mbox{\scriptsize s}}=4.24\times 10^{4}$ m${}^{-1}$
for
$v_{\mbox{\scriptsize g}}/c=0.05$.
This wavenumber corresponds to
$\omega_{\mbox{\scriptsize s}}=2.03\times 10^{15}$ s${}^{-1}$
and
$\lambda_{\mbox{\scriptsize s}}=9.29\times 10^{-7}$ m
in the dispersion relation shown in Fig.~\ref{Figure03ab} (b).
This wavelength is included in the range of that of transparent light for LiNbO${}_{3}$.
This signal light requires the pump light whose wavelength and angular frequency are equal to
$\lambda_{\mbox{\scriptsize p}}=4.65\times 10^{-7}$ m
and
$\omega_{\mbox{\scriptsize p}}(l_{\mbox{\scriptsize A}}+l_{\mbox{\scriptsize B}})/(2\pi c)=2.37$,
respectively,
and it is included in the eighth conduction band from the bottom as shown in Fig.~\ref{Figure03ab} (a).

Then, we calculate the
squeezing parameter $r=T^{(1)}\omega_{\mbox{\scriptsize s}}A\tilde{\chi}^{(2)}$.
A relationship between the intensity $I$ and an amplitude $A$ of the electric field of the pump laser is given by
\begin{equation}
I
=
\frac{1}{2}\epsilon_{0}cnA^{2},
\end{equation}
where $n$ is the refractive index of the vacuum.
We assume that the radiant flux and the radius of the laser beam are given by
$W=0.001$ W
and
$d=5.0\times 10^{-6}$ m, respectively.
(These are standard specifications of a semiconductor laser for commercial use.)
Because of
\begin{equation}
I
=
\frac{W}{\pi d^{2}},
\end{equation}
and the approximation $n\simeq 1$,
we get $A=9.79\times 10^{4}$ Vm${}^{-1}$.

Thus, from Eq.~(\ref{squeezing-parameter-r-1}),
we obtain the squeezing parameter as
$r=1.84$
where we use $\tilde{\chi}^{(2)}=25.2\times 10^{-12}$ mV${}^{-1}$.
This implies that the squeezing level $e^{2r}$ reaches $15.9$ dB.
As mentioned in Sec.~\ref{section-introduction},
an experimental demonstration of the generation of $15$ dB squeezed state was reported in \cite{Vahlbruch2016}.

Here, we consider a case where the frequency of the incident signal photon $\nu_{\mbox{\scriptsize s}}$ has an error $\Delta\nu_{\mbox{\scriptsize s}}$.
This inaccuracy gives rise to an error of the group velocity $v_{\mbox{\scriptsize g}}$.
However, we can cancel the error of $v_{\mbox{\scriptsize g}}$ by adjusting $N$,
that is to say,
adjusting the transparency of the partially reflective mirror in Fig.~\ref{Figure02}.
More specifically, by introducing the polarization of laser emission,
putting a polarization filter in front of the partially reflective mirror,
and rotating the axis of the polarization of the filter precisely,
we can change the actual transparency of the partially reflective mirror.
If we use the Pockels effect for the polarization filter,
we can control the direction of the polarization for the filtering by changing an applied voltage precisely.
Thus, we can avoid the trouble of high precision for $\nu_{\mbox{\scriptsize s}}$.

Second, we estimate the parameter of the displacement operator practically.
Letting the Hamiltonian $\hat{H}_{\mbox{\scriptsize II}}$ defined in Eq.~(\ref{Hamiltonian-II-displacement-operator-0}) cause the displacement operator $\hat{D}(\alpha/N)$
and using Eq.~(\ref{gamma-practical-2}),
we obtain
\begin{eqnarray}
\alpha
&=&
\Delta\tau_{2}\gamma N \nonumber \\
&=&
\frac{\sqrt{2}\pi c\Delta\tau_{2}}{\lambda_{\mbox{\scriptsize s}}}N
\frac{\sigma_{\mbox{\scriptsize em}}}{\sigma}\rho_{0}\Delta V.
\end{eqnarray}
Then, we consider a scenario where
$\lambda_{\mbox{\scriptsize s}}=9.29\times 10^{-7}$ m
and
$\Delta\tau_{2}=4\times 10^{-14}$ s.
This implies that the length of the gain medium for stimulated emission placed in the ring resonator is given by
$l_{\mbox{\scriptsize II}}=\Delta\tau_{2}c=1.20\times 10^{-5}$ m.
In general, the stimulated emission cross-section of the excited atom in the gain medium is given by
$\sigma_{\mbox{\scriptsize em}}\sim 10^{-24}$ m${}^{2}$
\cite{Koch1997,Philipps2001,Brown2021}.
For example, setting the density of the excited atoms in the gain medium
$\rho_{0}=1.45\times 10^{24}$ m${}^{-3}$,
and letting $\sigma=\pi d^{2}$ and $\Delta V=c\Delta\tau_{2}\sigma$,
we get $\alpha=1.00$.
The density $\rho_{0}$ is proportional to a current density input into the gain medium.
Thus, we can adjust $\rho_{0}$ by changing the value of the current.
According to \cite{Brown2021},
a density of Yb in Yb-doped phosphate QX laser glass is estimated at $1.51\times 10^{27}$ ions/m${}^{3}$.
Thus, our evaluation of $\rho_{0}$ is reasonable.

From Eqs.~(\ref{squeezed-coherent-state-emitted-photonic-crystal-1}), (\ref{squeezed-coherent-state-Lie-Trotter-1}),
and (\ref{squeezed-coherent-state-Lie-Trotter-2}),
letting the coherent squeezed like state generated by the ring resonator be given by
$e^{i\Theta}|-r,\alpha'\rangle$
and setting
$\alpha=1.00$, $r=1.84$, and $\varphi=\pi$,
we obtain
\begin{eqnarray}
\alpha'
&=&
\alpha
\Bigl[
\frac{\sinh r}{r}
+
e^{i\varphi}\frac{1}{r}(-1+\cosh r)
\Bigr] \nonumber \\
&=&
0.458.
\end{eqnarray}
Now, we calculate the radiant flux of the generated coherent squeezed like light.
An expectation value of the number of the photons of $|-r,\alpha'\rangle$ is given by
\begin{eqnarray}
&&
\langle -r,\alpha'|
\hat{a}^{\dagger}\hat{a}
|-r,\alpha'\rangle \nonumber \\
&=&
\langle\alpha'|
\hat{S}^{\dagger}(-r)\hat{a}^{\dagger}\hat{S}(-r)
\hat{S}^{\dagger}(-r)\hat{a}\hat{S}(-r)
|\alpha'\rangle \nonumber \\
&=&
\langle\alpha'|\hat{b}^{\dagger}\hat{b}|\alpha'\rangle \nonumber \\
&=&
\alpha'^{2}(\cosh^{2}r
+
2\sinh r\cosh r) \nonumber \\
&&
+
(1+\alpha'^{2})\sinh^{2}r,
\end{eqnarray}
where we use
\begin{equation}
\hat{b}
=
\hat{a}\cosh r+\hat{a}^{\dagger}\sinh r.
\end{equation}

Describing the amplitude of $|-r,\alpha'\rangle$ as $A$,
we obtain
\begin{eqnarray}
A^{2}
&=&
\frac{\hbar\omega_{\mbox{\scriptsize s}}}{2\epsilon_{0}V}
[
\alpha'^{2}(\cosh^{2}r
+
2\sinh r\cosh r) \nonumber \\
&&
+
(1+\alpha'^{2})\sinh^{2}r
].
\end{eqnarray}
Thus, the radiant flux $W$ of the coherent squeezed like light is estimated at
\begin{eqnarray}
W
&=&
I\pi d^{2} \nonumber \\
&=&
\frac{1}{2}\epsilon_{0}cA^{2}\pi d^{2} \nonumber \\
&=&
\frac{\hbar\omega_{\mbox{\scriptsize s}}}{4N(\Delta\tau_{1}+\Delta\tau_{2})}
[
\alpha'^{2}(\cosh^{2}r
+
2\sinh r\cosh r) \nonumber \\
&&
+
(1+\alpha'^{2})\sinh^{2}r
] \nonumber \\
&=&
\mbox{$2.31\times 10^{-9}$ W},
\end{eqnarray}
where we set the volume of quantization
$V=\pi d^{2}Nc(\Delta\tau_{1}+\Delta\tau_{2})$.

\section{\label{section-entanglement-squeezed-state}Entanglement generated by the squeezed light}
\begin{figure}
\begin{center}
\includegraphics[scale=1.0]{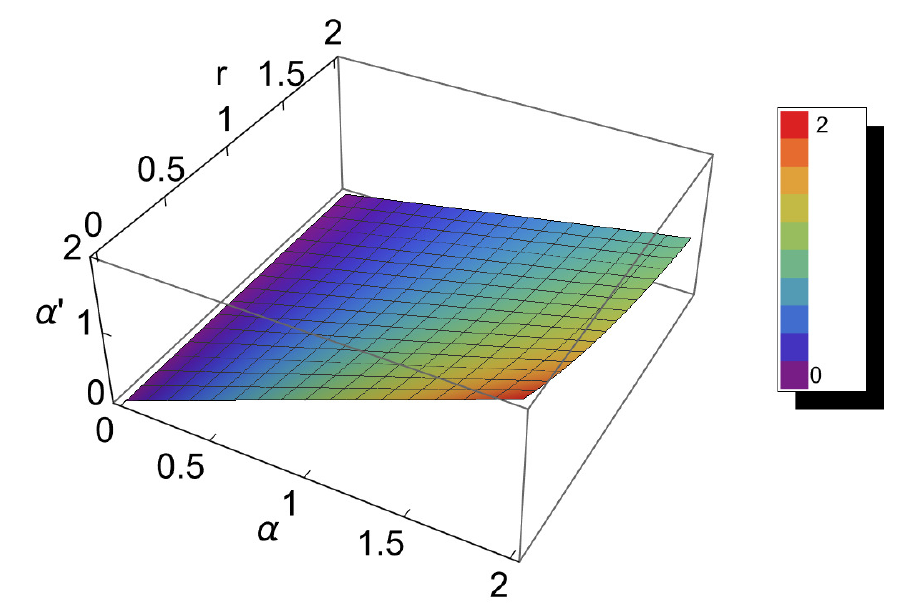}
\end{center}
\caption{A 3D plot of $\alpha'$ defined in Eq.~(\ref{alpha-dash-1}) as a function of $r$ and $\alpha$.
Looking at this plot, we note that $\alpha'$ decreases as $r$ gets larger and increases as $\alpha$ gets larger.}
\label{Figure05}
\end{figure}

When we split the coherent squeezed like light emitted from the ring resonator shown in Fig.~\ref{Figure02} into a pair of light beams with a 50-50 beam splitter,
these two beams have entanglement.
In this section, we examine this entanglement quantitatively.
Expressing the parameters of the displacement and squeezing operators of the ring resonator as $\alpha$ and $r$
and letting the state of the light emitted from the ring resonator be given by
\\
$|||-r,\alpha\rangle\!\rangle\!\rangle
=|-r,\alpha'\rangle$,
we obtain
\begin{equation}
\alpha'
=
\alpha
\Bigl[
\frac{\sinh r}{r}+e^{i\pi}\frac{1}{r}(-1+\cosh r)
\Bigr],
\label{alpha-dash-1}
\end{equation}
where we assume that $\alpha$ and $r$ are real.
Because $\alpha$ is real, $\Theta$ given by Eq.~(\ref{squeezed-coherent-state-Lie-Trotter-2}) is equal to zero.
In Fig.~\ref{Figure05}, we plot $\alpha'$ as a function of $r$ and $\alpha$.
Looking at this 3D graph, we note that $\alpha'$ decreases as $r$ gets larger and increases as $\alpha$ gets larger.

\begin{figure}
\begin{center}
\includegraphics[scale=0.7]{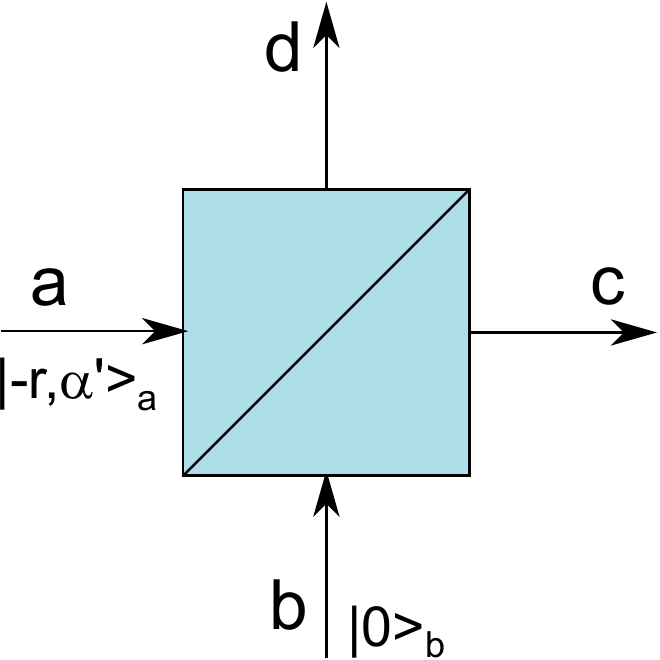}
\end{center}
\caption{Schematic of the 50-50 beam splitter.
We input $|-r,\alpha'\rangle_{a}$ and $|0\rangle_{b}$ into the modes $a$ and $b$, respectively.}
\label{Figure06}
\end{figure}

Next, we split $|-r,\alpha'\rangle$ into two light beams with the 50-50 beam splitter.
We describe these two beams as modes $a$ and $b$ as shown in Fig.~\ref{Figure06}.
To estimate entanglement between modes $a$ and $b$ quantitatively,
the following physical value is proposed
\cite{Duan2000,Simon2000}:
\begin{equation}
\mbox{criterion}
=
\langle\Delta^{2}(\hat{x}_{a}-\hat{x}_{b})\rangle
+
\langle\Delta^{2}(\hat{p}_{a}+\hat{p}_{b})\rangle,
\label{definition-criterion-1}
\end{equation}
\begin{eqnarray}
\hat{x}_{a}
=
\hat{a}^{\dagger}
+
\hat{a},
&\quad&
\hat{x}_{b}
=
\hat{b}^{\dagger}
+
\hat{b}, \nonumber \\
\hat{p}_{a}
=
i(\hat{a}^{\dagger}
-
\hat{a}),
&\quad&
\hat{p}_{b}
=
i(\hat{b}^{\dagger}
-
\hat{b}),
\end{eqnarray}
where $\hat{a}$ and $\hat{a}^{\dagger}$ denote annihilation and creation operators of the mode $a$,
and
$\hat{b}$ and $\hat{b}^{\dagger}$ denote annihilation and creation operators of the mode $b$, respectively.

The 50-50 beam splitter transforms the annihilation operators of the modes $a$ and $b$ as follows:
\begin{equation}
\hat{a}\to\hat{B}\hat{a}\hat{B}^{\dagger},
\quad
\hat{b}\to\hat{B}\hat{b}\hat{B}^{\dagger},
\end{equation}
\begin{equation}
\hat{B}
=
\exp[\frac{\pi}{4}(\hat{a}^{\dagger}\hat{b}-\hat{a}\hat{b}^{\dagger})].
\end{equation}
Thus, inputting $|-r,\alpha'\rangle_{a}$ and $|0\rangle_{b}$ into the beam splitter,
we obtain $\hat{B}^{\dagger}|-r,\alpha'\rangle_{a}|0\rangle_{b}$.
According to \cite{Kim2002,Azuma2022},
we can rewrite this state as
\begin{eqnarray}
&&
\hat{B}^{\dagger}|-r,\alpha'\rangle_{a}|0\rangle_{b} \nonumber \\
&=&
\hat{S}_{ab}(-\frac{r}{2})\hat{S}_{a}(-\frac{r}{2})\hat{S}_{b}(-\frac{r}{2})\hat{D}_{a}(\frac{\alpha'}{\sqrt{2}})\hat{D}_{b}(\frac{\alpha'}{\sqrt{2}}) \nonumber \\
&&
\times
|0\rangle_{a}|0\rangle_{b},
\end{eqnarray}
where
modes $a$ and $b$ on the right-hand side of the equation correspond to ports $c$ and $d$ in Fig.~\ref{Figure06}, respectively,
and
\begin{equation}
\hat{S}_{ab}(-\frac{r}{2})
=
\exp[\frac{r}{2}(\hat{a}^{\dagger}\hat{b}^{\dagger}-\hat{b}\hat{a})],
\end{equation}
\begin{eqnarray}
\hat{S}_{a}(-\frac{r}{2})
&=&
\exp[\frac{r}{2}(\hat{a}^{\dagger 2}-\hat{a}^{2})], \nonumber \\
\hat{S}_{b}(-\frac{r}{2})
&=&
\exp[\frac{r}{2}(\hat{b}^{\dagger 2}-\hat{b}^{2})],
\end{eqnarray}
\begin{eqnarray}
\hat{D}_{a}(\frac{\alpha}{\sqrt{2}})
&=&
\exp[\frac{\alpha}{\sqrt{2}}(\hat{a}^{\dagger}-\hat{a})], \nonumber \\
\hat{D}_{b}(\frac{\alpha}{\sqrt{2}})
&=&
\exp[\frac{\alpha}{\sqrt{2}}(\hat{b}^{\dagger}-\hat{b})].
\end{eqnarray}

Here, we want to obtain an explicit mathematically closed form of
$\hat{B}^{\dagger}|-r,\alpha\rangle_{a}|0\rangle_{b}$.
To accomplish it, we utilize the following convenient formula given by Eq.~(7.80) of \cite{Gerry2005}:
\begin{eqnarray}
||\zeta,\alpha\rangle\!\rangle
&=&
\frac{1}{\sqrt{\cosh r}}
\exp
\Bigl[
-\frac{1}{2}|\alpha|^{2}
-\frac{1}{2}\alpha^{*2}e^{i\theta}\tanh r
\Bigr] \nonumber \\
&&
\times
\sum_{n=0}^{\infty}
\frac{[(1/2)e^{i\theta}\tanh r]^{n/2}}{\sqrt{n!}} \nonumber \\
&&
\times
H_{n}[\xi(e^{i\theta}\sinh (2r))^{-1/2}]|n\rangle,
\label{squeezed-coherent-state-zeta-alpha-formula-1}
\end{eqnarray}
where $\alpha$ is complex,
\begin{equation}
\xi
=
\alpha \cosh r+\alpha^{*}e^{i\theta}\sinh r,
\end{equation}
$H_{n}(x)$ represents the Hermite polynomial,
and
$\zeta=r e^{i\theta}$.
Using Eq.~(\ref{conversion-squeezed-coherent-state-1}) which transforms $||\zeta,\alpha\rangle\!\rangle$ into $|\zeta,\alpha\rangle$,
we rewrite Eq.~(\ref{squeezed-coherent-state-zeta-alpha-formula-1}) as the following mathematical expression
that is suitable for numerical calculations:
\begin{eqnarray}
CS_{n}^{a}
&=&
{}_{a}\langle n|-\frac{r}{2},\frac{\alpha'}{\sqrt{2}}\rangle_{a} \nonumber \\
&=&
\frac{1}{\sqrt{\cosh (r/2)}}
\exp
\Bigl[
-\frac{1}{2}\tilde{\alpha}^{2}(\frac{\alpha'}{\sqrt{2}},-\frac{r}{2})(1-\tanh (\frac{r}{2}))
\Bigr] \nonumber \\
&&
\times
\frac{1}{\sqrt{n!}}
i^{n}
[\frac{1}{2}\tanh(\frac{r}{2})]^{n/2} \nonumber \\
&&
\times
H_{n}(\tilde{\xi}(\tilde{\alpha}(\frac{\alpha'}{\sqrt{2}},-\frac{r}{2}),-\frac{r}{2})(-i)\sinh^{-1/2}r),
\end{eqnarray}
where $\alpha'$ is real and
\begin{eqnarray}
\tilde{\alpha}(\alpha,r)
&=&
\alpha(\cosh r-\sinh r), \nonumber \\
\tilde{\xi}(\alpha,r)
&=&
\alpha(\cosh r+\sinh r).
\end{eqnarray}
Moreover, we use the following formula
\cite{Azuma2022}:
\begin{eqnarray}
S^{ab}_{n_{1}n_{2},lk}
&=&
{}_{a}\langle n_{1}|{}_{b}\langle n_{2}|
\hat{S}_{ab}(-\frac{1}{2}r)
|l\rangle_{a}|k\rangle_{b} \nonumber \\
&=&
\sum_{n = 0}^{\mbox{\scriptsize min}[n_{1},n_{2}]}
\sum_{m = 0}^{\mbox{\scriptsize min}[l,k]}
\delta_{l-m,n_{1}-n}
\delta_{k-m,n_{2}-n}
(-1)^{m}
\frac{1}{m!n!} \nonumber \\
&&
\times
\tanh^{m+n}(\frac{r}{2}) \nonumber \\
&&
\times
\exp
\Biggl\{
-(l+k-2m+1)\ln[\cosh(\frac{r}{2})]
\Biggr\} \nonumber \\
&&
\times
\frac{\sqrt{l!k!n_{1}!n_{2}!}}{(l-m)!(k-m)!}.
\end{eqnarray}
From the above formulae, we obtain matrix elements in the form,
\begin{equation}
{}_{a}\langle n_{1}|{}_{b}\langle n_{2}|\hat{B}^{\dagger}|-r,\alpha'\rangle_{a}|0\rangle_{b}
=
\sum_{l=0}^{\infty}\sum_{k=0}^{\infty}
S^{ab}_{n_{1} n_{2}, l k}
CS_{l}^{a}CS_{k}^{b}.
\label{n1n2-B-minus-r-alphadash-0}
\end{equation}

Now, we calculate the criterion defined in Eq.~(\ref{definition-criterion-1}).
Obviously,
$\langle\hat{x}_{a}\rangle=\langle\hat{x}_{b}\rangle$,
$\langle\hat{x}_{a}^{2}\rangle=\langle\hat{x}_{b}^{2}\rangle$,
$\langle\hat{p}_{a}\rangle=\langle\hat{p}_{b}\rangle$,
and
$\langle\hat{p}_{a}^{2}\rangle=\langle\hat{p}_{b}^{2}\rangle$
hold.
Thus, we obtain
\begin{eqnarray}
\mbox{criterion}
&=&
8\langle\hat{a}^{\dagger}\hat{a}\rangle
+
4
-
8\mbox{Re}[\langle\hat{a}^{\dagger}\hat{b}^{\dagger}\rangle] \nonumber \\
&&
+
8(\mbox{Re}[\langle\hat{a}^{\dagger}\rangle^{2}]-|\langle\hat{a}^{\dagger}\rangle|^{2}).
\label{criterion-2}
\end{eqnarray}
We can compute
$\langle\hat{a}^{\dagger}\rangle$,
$\langle\hat{a}^{\dagger}\hat{a}\rangle$, and $\langle\hat{a}^{\dagger}\hat{b}^{\dagger}\rangle$
as follows:
\begin{eqnarray}
\langle\hat{a}^{\dagger}\rangle
&=&
\sum_{n_{1}=0}^{\infty}
\sum_{n_{2}=0}^{\infty}
{}_{a}\langle -r,\alpha'|{}_{b}\langle 0|\hat{B}|n_{1}+1\rangle_{a}|n_{2}\rangle_{b} \nonumber \\
&&
\times
\sqrt{n_{1}+1} \nonumber \\
&&
\times
{}_{a}\langle n_{1}|{}_{b}\langle n_{2}|\hat{B}^{\dagger}|-r,\alpha'\rangle_{a}|0\rangle_{b},
\label{mean-value-a-dagger}
\end{eqnarray}
\begin{equation}
\langle\hat{a}^{\dagger}\hat{a}\rangle
=
\sum_{n_{1}=0}^{\infty}
\sum_{n_{2}=0}^{\infty}
n_{1}
|{}_{a}\langle n_{1}|{}_{b}\langle n_{2}|\hat{B}^{\dagger}|-r,\alpha'\rangle_{a}|0\rangle_{b}|^{2},
\label{mean-value-a-dagger-a}
\end{equation}
\begin{eqnarray}
\langle\hat{a}^{\dagger}\hat{b}^{\dagger}\rangle
&=&
\sum_{n_{1}=0}^{\infty}
\sum_{n_{2}=0}^{\infty}
{}_{a}\langle -r,\alpha'|{}_{b}\langle 0|\hat{B}|n_{1}+1\rangle_{a}|n_{2}+1\rangle_{b} \nonumber \\
&&
\times
\sqrt{(n_{1}+1)(n_{2}+1)} \nonumber \\
&&
\times
{}_{a}\langle n_{1}|{}_{b}\langle n_{2}|\hat{B}^{\dagger}|-r,\alpha'\rangle_{a}|0\rangle_{b}.
\label{mean-value-a-dagger-b-dagger}
\end{eqnarray}
From these results, we can express the criterion as a function of $r$ and $\alpha'$
that is suitable for numerical calculations.

\begin{figure}
\begin{center} 
\includegraphics[scale=0.85]{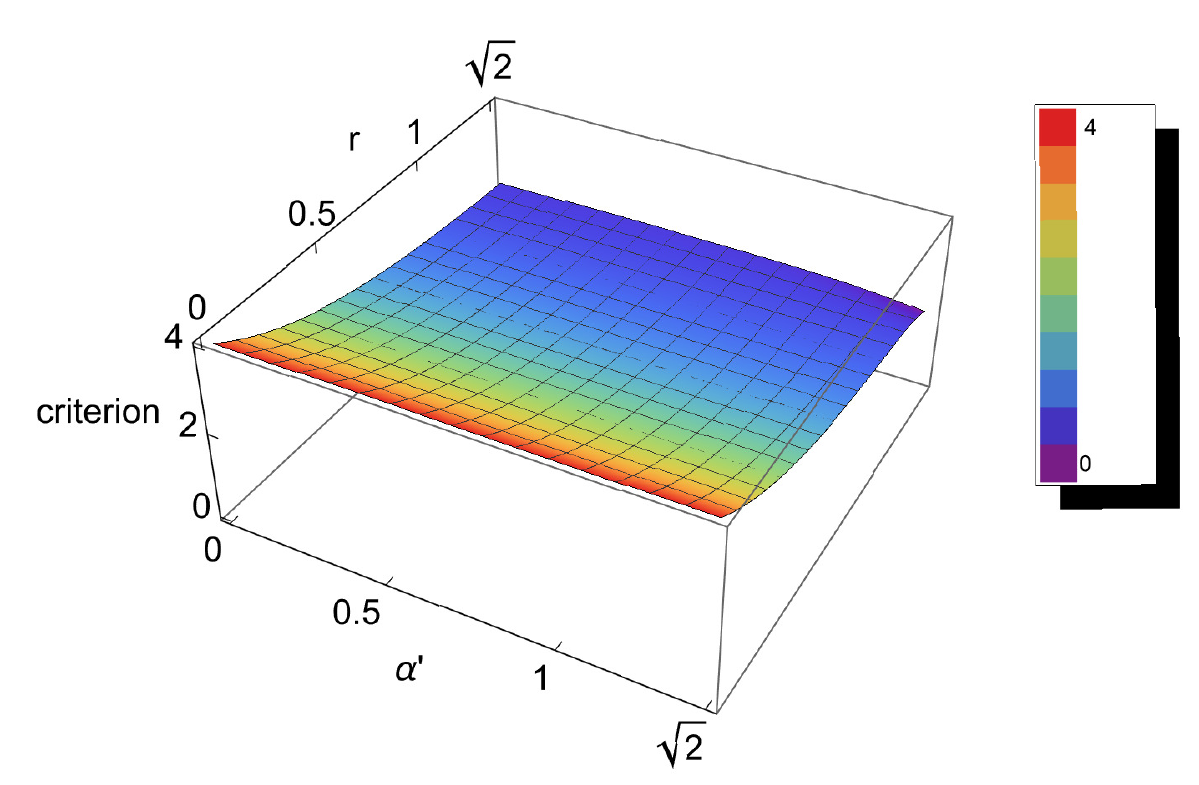}
\end{center}
\caption{A 3D plot of the criterion defined in Eq.~(\ref{definition-criterion-1}) as a function of $r$ and $\alpha'$.
If $r>0$, the criterion is less than four and the state of the modes $a$ and $b$ becomes entangled.
To compute Eqs.~(\ref{n1n2-B-minus-r-alphadash-0}), (\ref{mean-value-a-dagger}), (\ref{mean-value-a-dagger-a}),
and (\ref{mean-value-a-dagger-b-dagger}) numerically,
we carry out summations from zero to $60$ instead of infinity.}
\label{Figure07}
\end{figure}

We show numerical calculations of the criterion in Fig.~\ref{Figure07}.
When $r=0$ holds, the criterion is equal to four regardless of the value of $\alpha'$.
The necessary and sufficient condition for the separability of Gaussian states of two modes is $\mbox{criterion}\geq 4$.
(The Gaussian state is a state whose Wigner function is Gaussian.)
Thus, if $r>0$, the criterion is less than four and the two-mode state is inseparable.

\section{\label{section-discussions}Discussions}
In this paper, we propose a method to generate coherent squeezed like states with large squeezing levels using the ring resonator
where the one-dimensional nonlinear photonic crystal and the gain medium for stimulated emission are placed.
Moreover, we split the coherent squeezed like light with the 50-50 beam splitter and obtain the two entangled light beams.
We compute their entanglement quantitatively.
Giving practical parameters of physical quantities of the ring resonator, the signal light, and the pump light,
we confirm that we can execute our proposed method experimentally.

Looking at Fig.~\ref{Figure07}, we note that the entanglement increases as $r$ gets larger than zero.
Thus, we can conclude that we can generate entangled photons efficiently from a two-mode squeezed state using the beam splitter.
Carrying out numerical calculations for Fig.~\ref{Figure07},
we find that we can hardly obtain stable numerical results of the criterion for $r>\sqrt{2}$ or $\alpha'>\sqrt{2}$.
The reason why this trouble happens is as follows.
Equations~(\ref{n1n2-B-minus-r-alphadash-0}), (\ref{mean-value-a-dagger}), (\ref{mean-value-a-dagger-a}),
and (\ref{mean-value-a-dagger-b-dagger})
include summations from zero to infinity.
In actual numerical computations,
we must replace these infinite summations with finite ones.
This approximation makes the criterion not converge to a specific value for $r>\sqrt{2}$ or $\alpha'>\sqrt{2}$.
Therefore, improvement of the method for numerical estimation of the criterion is our future challenge.

\ack
This work was supported by MEXT Quantum Leap Flagship Program Grant No. JPMXS0120351339.
\\

\end{document}